\documentclass[aps,prl,twocolumn,showpacs,preprintnumbers]{revtex4-1}

\usepackage{psfrag,graphicx}
\usepackage{dcolumn}
\usepackage{amsmath,amssymb}
\usepackage{bm}
\usepackage{amsfonts,amssymb,amsmath}        
\usepackage{epstopdf}

\newcommand{\be}{\begin{equation}}
\newcommand{\ee}{\end{equation}}
\newcommand{\bq}{\begin{eqnarray}}
\newcommand{\eq}{\end{eqnarray}}

\begin{document}
\title{Getting the public involved in Quantum Error Correction}
\author{James R. Wootton}
\affiliation{Department of Physics, University of Basel, Klingelbergstrasse 82, CH-4056 Basel, Switzerland}

\begin{abstract}

The Decodoku project seeks to let users get hands-on with cutting-edge quantum research through a set of simple puzzle games. The design of these games is explicitly based on the problem of decoding qudit variants of surface codes. This problem is presented such that it can be tackled by players with no prior knowledge of quantum information theory, or any other high-level physics or mathematics. Methods devised by the players to solve the puzzles can then directly be incorporated into decoding algorithms for quantum computation. In this paper we give a brief overview of the novel decoding methods devised by players, and provide short postmortem for Decodoku v1.0-v4.1.

\end{abstract}

\maketitle

\section{Introduction}

\begin{figure}[t]
\begin{center}
{\includegraphics[width=\columnwidth]{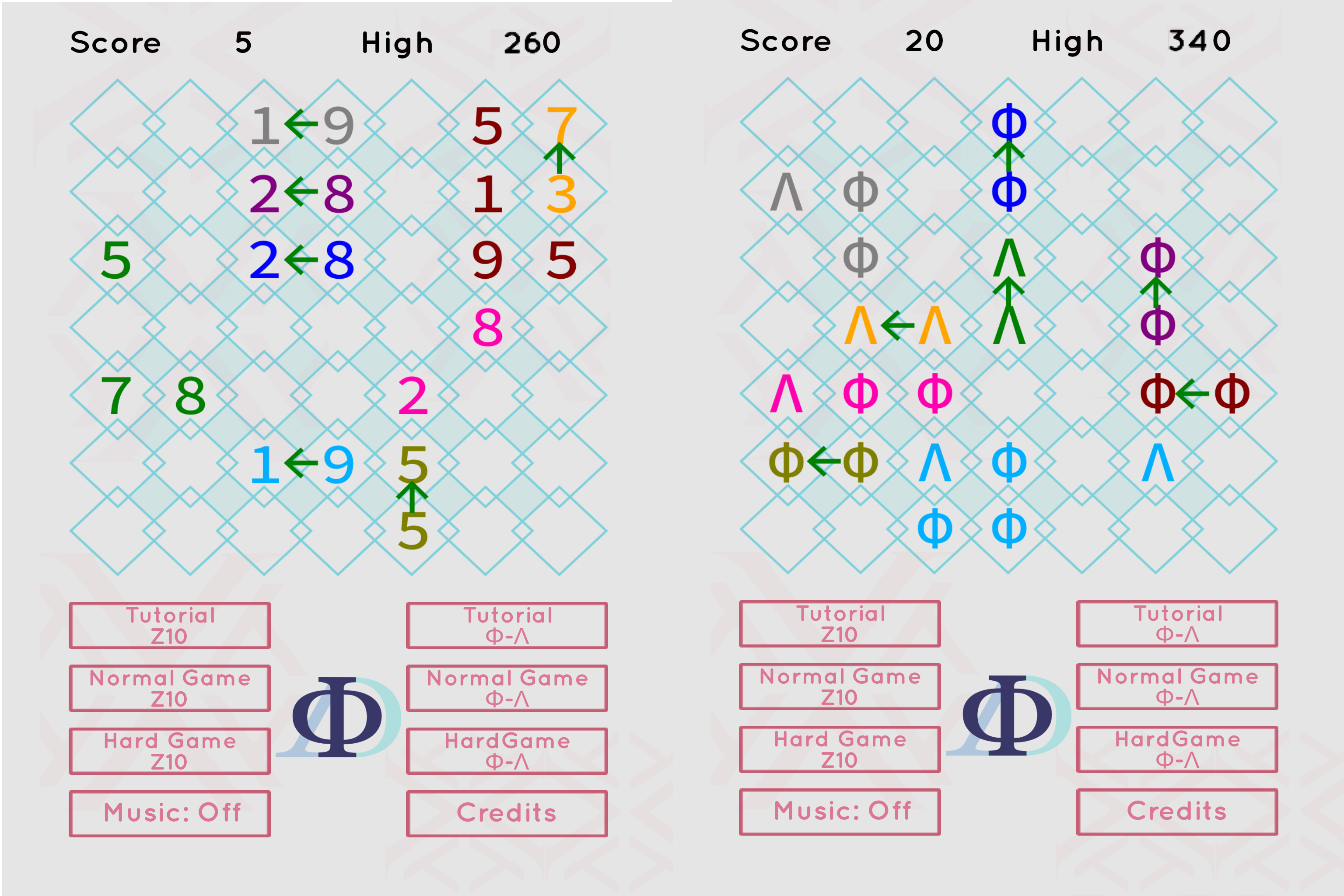}}
\caption{\label{screenshots} (Left) Screenshot of Tutorial mode for a Z10 game of Decodoku. (Right) Screenshot of Tutorial mode for a $\Phi$-$\Lambda$ game of Decodoku. In both cases, suggested moves are shown. These are calculated by a method that is based on those developed by players.
}
\end{center}
\end{figure}

The Decodoku games are based on the grid shown in the screenshots of Fig. \ref{screenshots}. These have an array of large interlocking squares (known as \emph{plaquettes}), with small squares forming at their intersections. These small squares represent \emph{qudits}, physical systems that will form the basic building of quantum computers~\cite{nielsen:00}.

The purpose of the grid layout is to implement a surface code \cite{double,dennis}, which allows errors on qudits to be detected and corrected. This is done by computing a quantity for each plaquette that depends on the states of all the qudits that surround it. Certain values of these quantities serve as signatures that errors have occurred. By looking at what the values are, and on which plaquettes they occur, a strategy to correct the errors can be determined.

This process, known as decoding, can be compared to solving a puzzle. For codes based on \emph{qubits}, the simplest form of qudit, highly effective algorithms for decoding have been found~\cite{dennis}. However, current decoding methods for more complex qudits are quite heuristic~\cite{hutter:improved}. Finding more effective methods is therefore a perfect candidate for a citizen science project. The puzzles at the heart of decoding simply need to be presented to players, and insights into quantum error correction can be extracted from their solutions. This is the aim of Decodoku.

Full details on how the Decodoku games are defined, and how they relate to quantum error correction, can be found at \cite{decodoku:blog}.

\section{Postmortem for v1.0-v4.1}

It is common in game development to create a `postmortem' for games whose development has finished. This allows reflection on the lessons learned during the process. Here we provide a short postmortem for the first phase of Decodoku development, which terminated at version 4.1.

The Decodoku project has two main aims. The primary aim is to give players a taste of what it is like to be a scientist. We aim to show them the kind of problems tackled by researchers working in the field of quantum error correction, and how to go about solving them. The game-based nature of the endeavour makes this experience accessible even to those without significant scientific or mathematical knowledge. The secondary aim of the project is to allow the players to actually contribute to science.

For the most part we expect users to be casual gamers. They will download Decodoku (and its sister games, Decodoku:Puzzles and Decodoku:Colors) knowing that there is a scientific background based in quantum technologies.

By playing the games, they should hopefully realize that this science is not so mysterious as they might previously have imagined, and will then engage more readily with issues relating to quantum technologies in future. Such a result is regarded as meeting the primary aim of the project, even if the player chooses not to engage further.

A fraction of players will go beyond this. They will seek to learn obtain the highest score and compare their results and methods with other players. It is those for which the secondary aim of the project is relevant.

In current versions of Decodoku, no data regarding game play is gathered. This is unlike other quantum citizen science games, such as \emph{Quantum Moves} \cite{quantum_moves}, where players generate data to be analysed by professional scientists. In Decodoku it is up to the players themselves to analyse their method and generate a scientific result.

This may sound like a difficult burden for the players, but it is nothing beyond what often seen in player behaviour. For any suffiently popular game, a subset of players will systematically strive to find an optimal strategy. They share their insights, evaluate the results of others and build upon them. This parallels the problem solving, dissemination, peer review and overall `social knowledge construction' \cite{ref3} process of professional scientists. For an example, an analysis of forum disucssions for `World of Warcraft' was made in \cite{ref4}.

For Decodoku, interested players were encouraged to discuss their methods on the project's messageboard \cite{decodoku:subreddit}. Within the first few months it was found that the favoured methods of dissemination were primarily text-based. Gameplay videos were rare and were always devoid of commentary, other than explanatory text. In response we updated the game (v4.0) to make it easier to provide good text-based explanations. The updated game created save files, which provided a text-based record of all moves, and also allowed players to make annotations.

Through this method, we recieved a number of high quality results. The insights they have brought are summarized in the next section. The secondary aim of the project was therefore certainly fulfilled.

Even so, the number of such results was small. Though this will be partly due to the limited userbase of the games, we should also assume that more results would have been gathered if users found it easier to contribute.

The solution to this problem is not obvious. The nature of the scientific problem means that players must reflect upon and explain their strategies in order to contribute. This cannot be replaced with any in-game automated process. Furthermore, any attempt at gathering such information in-game should also be careful not to leave casual gamers with the impression that the science is demanding and techical, since this would reduce the success of the project's primary aim.

In the current version (v5.0) some efforts have been made to address this issue. This includes an interactive tutorial which suggests possible moves to the player. It is hoped that this will make the game more accessible for casual gamers, but also provide potential citizen scientists with a reference point when analysing their method. For example they could look at which suggested moves they do and do not use. They can also try to determine what kinds of move they apply that are never suggested, and so get an insight into how their method expands upon existing knowledge.

\section{Decoding based on player methods}

The philosophy of the Decodoku project is that the players themselves analyse and present their decoding methods. Their discoveries can then be cited directly when built upon by scientists in the field. This differs from other citizen science projects in which the data is collected, analysed and published by the scientists behind the project.  The information submitted by players is available at \cite{decodoku:results}. Here we summarize the results.

In this section, non-trivial syndrome elements are referred to as \emph{defects}. In the context of the games, defects are represented by plaquettes containing a number or other symbol. We also use \emph{cluster} to refer to the independent groups of defects formed when errors affect an overlapping set of plaquettes. In Tutorial Mode for Decodoku, these clusters are distinguished using different colours. In Decodoku:Colors, they are distinguished by using both different colours and symbols.

\subsection{Decodoku and Decodoku:Colors}

The majority of methods submitted by players pertained to the $Z10$ variant of Decodoku, which represents the decoding problem for a surface code of 10 level qudits. In the game, new errors arise as the player attempts to correct previous ones. The rate at which new errors occur exceeds the rate of correction, and so the player must struggle to keep the errors under control as long as they can.

In a tutorial mode, and in the simplified game \emph{Decodoku:Colors}, the information regarding the clustering of defects is provided to the player. This allows them to concentrate on how to prioritize the correction of errors.

Player methods were then used to develop a computer controlled player (the \emph{bot}) for this Decodoku tutorial. This was mostly based on the results of the highest scoring player: aesche1988~\cite{aesche1988} This incorporates the main elements of methods provided by players, and can provide a guide for new ones to learn from. It also provides a framework in which fuller analysis of player data can be performed.

The bot requires the following quantities to be computed for each defect.

\begin{itemize}

\item Number of defects in the same cluster, including the defect considered.

\item Position of the center of the defect’s cluster. This is calculated as the centroid, using the Manhattan distance metric.

\end{itemize}

All possible pairs of defects are also considered, and the following quantities are computed.

\begin{itemize}

\item Manhattan distance between the defects.

\item Whether or not they belong to the same cluster.

\item Distance between the defects via he centre of the group. This is the sum of the distances to the centre of the group for both the defects in the pair. Note that this will be very high if both are far from the center, even if they are very close.

\item Whether or not they annihilate.

\item If the pair itself does not annihilate, it could be that adding a neighbour of one of the defects would form a triplet that annihilates. The number of such \emph{helpful neighbours} is found.

\item For each defect in the pair, the total number of neighbours is found. This does not include the other element of the pair if this is a neighbour.

\end{itemize}

Given all this information, the bot assigns a ranking of pairs. The bot then moves together the defects of the best ranked pair. In order to reproduce some of the behaviour of player methods, the ranking is defined by the following rules. In this, earlier rules have precedence over later ones.

\begin{enumerate}

\item Pairs that are part of the same group are ranked higher than those that aren’t.

\item Pairs separated by a distance less than 4 are ranked higher than those separated by higher distances. The smaller the distance, the better. This prioritizes pairs that are very close.

\item For pairs with a distance of 4 or greater, larger distances are ranked higher. This ensures we do not forget to bring in defects that have been moved far away by large error structures.

\item Pairs for with a larger distance via the centre of the group are ranked higher. This ensures that, if all else is equal, we focus on pairs that are both fairly far from the centre of their group.

\item If all else is equal, pairs for which only one has neighbours are ranked highest, followed by pairs for which neither have neighbours, followed by pairs for which both have neighbours. This prioritizes pruning in defects that are escaping from clusters, instead of getting rid of isolated pairs. Pairs in the middle of a big patch of numbers are dealt with last.

\end{enumerate}

The bot is found to not be as effective as the player methods on which it was based. Improvement could be provided through analysis of player data, and using it to determine better parameters for the ranking system. This will allow Decodoku to become a more traditional citizen science project, in which the main contribution of players is to create data for scientific analysis.

\subsection{Decodoku:Puzzles}

\emph{Decodoku:Puzzles} differs from \emph{Decodoku} and \emph{Decodoku:Colors} in that the puzzle is static: no new errors arise as the puzzle is completed. This version of the game was the focus of an impressive effort by one particular player. This contributor created a genetic algorithm designed to determine solutions to the decoding puzzles. Explanation of this method, as well as the source code itself, can be found at \cite{weasel}.

Given this method, an effort was made to find a deterministic algorithm to capture similar results. This was done through manual inspection of multiple instances of the genetic algorithm as it converged towards a solution. The deterministic algorithm found through this bears little similarity to the original genetic algorithm. However it does introduce a novel means of decoding of topological codes.

A basic outline of the decoder is as follows.

\begin{enumerate}

\item Each non-trivial syndrome element is labelled as an non-neutral cluster.

\item Each non-neutral cluster is merged with its nearest located cluster.

\item For each cluster, it is determined whether all elements of the cluster would annihilate if combined. Such clusters are relabelled as `neutral'.

\item Steps 2 and 3 are repeated until all clusters are neutral. The correction is then performed through the corresponding annihilations.

\end{enumerate}

This method has similarities to other HDRG decoders \cite{hutter:improved}. However, note that neutral clusters are not removed from the decoding procedure entirely. They no longer seek to merge with other clusters, but they can still be involved in a merge if they are the nearest neighbour of a non-neutral cluster. This addressed the same issue as the notion of \emph{shortcuts} in \cite{hutter:improved}, and may indeed be more effective in doing so. Specifically,  it could be that they are not susceptible to the fractal error structures fatal to other HDRG decoders \cite{dennis:thesis}. However, this remains to be determined. Some preliminary analysis of this method was applied in \cite{heim:thesis}.

\section{Conclusions}

In conclusion, the Decodoku project allows casual players a taste of the science behind quantum computing and the scientific method, and allows more involved players to truly work as scientists themselves. Players have provided interesting and useful results, which could form the basis of decoding procedures for qudit surface codes.

Despite this success, we have found that trying serve multiple types of player at once can be a great challenge. We hope that these experiences will be of use in the developement of similar citizen science projects in future.

\section{Acknowledgments}

The author ackowledges the NCCR QSIT for support.

\bibliography{refs}

\end{document}